\documentclass{article}
\usepackage{arxiv}
\usepackage[utf8]{inputenc} 
\usepackage[T1]{fontenc}    
\usepackage{hyperref}       
\usepackage{url}            
\usepackage{booktabs}       
\usepackage{amsfonts}       
\usepackage{nicefrac}       
\usepackage{microtype}      
\usepackage{lipsum}		
\usepackage{graphicx}
\usepackage{mathrsfs}
\usepackage{doi}
\usepackage{amsmath}
\usepackage{algorithm}
\usepackage{algorithmicx}
\usepackage[noend]{algpseudocode}
\usepackage{caption,subcaption}
\allowdisplaybreaks
\algnewcommand{\LeftComment}[1]{\Statex \(\triangleright\) #1}
\usepackage{stmaryrd}
\usepackage{color,soul}
\usepackage{wrapfig}
\usepackage{cite}
\hypersetup{
    colorlinks=true,
    linkcolor=blue,
    filecolor=magenta,      
    urlcolor=cyan,
}
\usepackage{courier}

\usepackage{listings}
\usepackage{xcolor}
\definecolor{codegreen}{rgb}{0,0.6,0}
\definecolor{codegray}{rgb}{0.5,0.5,0.5}
\definecolor{codepurple}{rgb}{0.58,0,0.82}
\definecolor{backcolour}{rgb}{0.95,0.95,0.92}

\lstdefinestyle{mystyle}{
  backgroundcolor=\color{backcolour},   commentstyle=\color{codegreen},
  keywordstyle=\color{magenta},
  numberstyle=\tiny\color{codegray},
  stringstyle=\color{codepurple},
  basicstyle=\ttfamily\footnotesize,
  breakatwhitespace=false,         
  breaklines=true,                 
  captionpos=b,                    
  keepspaces=true,                 
  numbers=left,                    
  numbersep=5pt,                  
  showspaces=false,                
  showstringspaces=false,
  showtabs=false,                  
  tabsize=2
}

\title{A simple framework for arriving at bounds on effective moduli in heterogeneous anisotropic poroelastic solids}

\author{
  {
  Saumik Dana}\\
	University of Southern California\\
	Los Angeles, CA 90007 \\
	\texttt{sdana@usc.edu} \\
}

\date{}

\begin{document}
\maketitle
\begin{abstract}
The concepts of representative volume element (RVE), statistical homogeneity and homogeneous boundary conditions are invoked to arrive at bounds on effective moduli for heterogeneous anisotropic poroelastic solids. The homogeneous displacement boundary condition applicable to linear elasticity is replaced by a homogeneous displacement-pressure boundary condition to arrive at an upper bound within the RVE while the homogeneous traction boundary condition applicable to linear elasticity is replaced by a homogeneous traction-fluid content boundary condition to arrive at a lower bound within the RVE. Statistical homogeneity is then invoked to argue that the bounds obtained over the RVE are representative of the bounds obtained over the whole heterogeneous poroelastic solid.
\end{abstract}
\section{Introduction}
The concepts of RVE~\cite{hashin-1962,hill-1963,hillmandel1,hill-1972,hashin-1983,zohdi,kachanov} commonly used in the analysis of composite materials, as well as statistical homogeneity~\cite{beran,hashinnasa,kroner} are invoked to analyse porous solids. This approach, as opposed to the theory of two-scale homogenization~\cite{suquet,bakhvalov,terada-2001,hornung,fishbook}, offers avenues for arriving at bounds on effective poroelastic moduli. These upscaled moduli can then be used to solve poromechanics on a coarser grid in the popular fixed stress split algorithm~\cite{dana-2018,dana2019design,dana2019system,dana2020,dana2021,danacg,danacmame,danathesis} for the solution of the coupled flow and poromechanics problem. RVE based methods decouple analysis of a composite material into analyses at the local and
global levels. The local level analysis models the microstructural details to determine effective
elastic properties by applying boundary conditions to the
RVE and solving the resultant boundary value problem. The composite structure is then
replaced by an equivalent homogeneous material having the calculated effective properties. The
global level analysis calculates the effective or average stress and strain within the equivalent
homogeneous structure. The applied boundary conditions,
however, cannot represent all the possible in-situ boundary conditions to which the RVE is
subjected within the composite. The accuracy of the RVE approximation depends on how well
the assumed boundary conditions reflect each of the myriad boundary conditions to which the
RVE is subjected in-situ. This is where the imposition of homogeneous boundary conditions plays a special role in arriving at bounds on effective properties. 
This paper is structured as follows: the model equations are presented in Section 2, the concept of homogeneous boundary conditions in linear elasticity are explained in Section 3, the procedural framework for arriving at bounds on effective poroelastic moduli are presented in Section 4 and the conclusions and outlook are presented in Section 5.
\section{Model equations}
\subsection{Flow model}\label{flowmodel1} 
Let the boundary $\partial \Omega=\Gamma_D^f \cup \Gamma_N^f$ where $\Gamma_D^f$ is the Dirichlet boundary and $\Gamma_N^f$ is the Neumann boundary. The fluid mass conservation equation \eqref{massagain1} in the presence of deformable and anisotropic porous medium with the Darcy law \eqref{darcyagain1} and linear pressure dependence of density \eqref{compressible1} with boundary conditions \eqref{bcagain11} and initial conditions \eqref{flowendagain1} is
\begin{align}
\label{massagain1}
&\frac{\partial \zeta}{\partial t}+\nabla \cdot \mathbf{z}=q\\
\label{darcyagain1}
&\mathbf{z}=-\frac{\mathbf{K}}{\mu}(\nabla p-\rho_0 \mathbf{g})=-\boldsymbol{\kappa}(\nabla p-\rho_0 \mathbf{g})\\
\label{compressible1}
&\rho=\rho_0(1+c\,(p-p_0))\\
\label{bcagain11}
&p=g \,\, \mathrm{on}\,\,\Gamma_D^f \times (0,T],\,\,\mathbf{z}\cdot\mathbf{n}=0 \,\, \mathrm{on}\,\,\Gamma_N^f \times (0,T]
\\
\label{flowendagain1}
&p(\mathbf{x},0)=p_0(\mathbf{x}),\,\,\rho(\mathbf{x},0)=\rho_0(\mathbf{x}),\,\, \phi(\mathbf{x},0)=\phi_0(\mathbf{x})\qquad (\forall \mathbf{x}\in \Omega)
\end{align}
where $p:\Omega \times (0,T]\rightarrow \mathbb{R}$ is the fluid pressure, $\mathbf{z}:\Omega \times (0,T]\rightarrow \mathbb{R}^3$ is the fluid flux, $\zeta$ is the fluid content, $\mathbf{n}$ is the unit outward normal on $\Gamma_N^f$, $q$ is the source or sink term, $\mathbf{K}$ is the uniformly symmetric positive definite absolute permeability tensor, $\mu$ is the fluid viscosity, $\rho_0$ is a reference density, $\phi$ is the porosity, $\boldsymbol{\kappa}=\frac{\mathbf{K}}{\mu}$ is a measure of the hydraulic conductivity of the pore fluid, $c$ is the fluid compressibility and $T>0$ is the time interval. 
\subsection{Poromechanics model}\label{poromodel1}
Let the boundary $\partial \Omega=\Gamma_D^p \cup \Gamma_N^p$ where $\Gamma_D^p$ is the Dirichlet boundary and $\Gamma_N^p$ is the Neumann boundary. Linear momentum balance for the anisotropic porous solid in the quasi-static limit of interest \eqref{mechstart} with small strain assumption \eqref{smallstrain} with boundary conditions \eqref{bc1} and initial condition \eqref{ic1} is
\begin{align}
\label{mechstart}
&\nabla\cdot \boldsymbol{\sigma}+\mathbf{f}=\mathbf{0}\\
\label{bforce}
&\mathbf{f}=\rho \phi\mathbf{g} + 
\rho_r(1-\phi)\mathbf{g}\\
\label{smallstrain}
&\boldsymbol{\epsilon}(\mathbf{u})=\frac{1}{2}(\nabla \mathbf{u} + (\nabla \mathbf{u})^T)\\
\label{bc1}
&\mathbf{u}\cdot\mathbf{n}_1=0\,\, \mathrm{on}\,\,\Gamma_D^p \times [0,T],\,\,
\boldsymbol{\sigma}^T\mathbf{n}_2=\mathbf{t}\,\, \mathrm{on}\,\,\Gamma_N^p \times [0,T]\\
\label{ic1}
&\mathbf{u}(\mathbf{x},0)=\mathbf{0}\qquad \forall \mathbf{x}\in \Omega
\end{align}
where $\mathbf{u}:\Omega \times [0,T]\rightarrow \mathbb{R}^3$ is the solid displacement, $\rho_r$ is the rock density, $\mathbf{f}$ is the body force per unit volume, $\mathbf{n}_1$ is the unit outward normal to $\Gamma_D^p$,  $\mathbf{n}_2$ is the unit outward normal to $\Gamma_N^p$, $\mathbf{t}$ is the traction specified on $\Gamma_N^p$, $\boldsymbol{\epsilon}$ is the strain tensor, $\boldsymbol{\sigma}$ is the Cauchy stress tensor given by the generalized Hooke's law~\cite{coussy}
\begin{align}
\label{constitutive}
\boldsymbol{\sigma}=\mathbb{M}\boldsymbol{\epsilon}
-\boldsymbol{\alpha} p=\boldsymbol{\sigma}'
-\boldsymbol{\alpha} p
\end{align}
where $\boldsymbol{\sigma}'$ is the effective stress, $\mathbb{M}$ is the fourth order anisotropic elasticity tensor and $\boldsymbol{\alpha}$ is the Biot tensor. 
The symmetry of the stress and strain tensor shows that
\begin{align*}
\tag{$i,j,k,l=1,2,3$}
&\mathbb{M}_{ijkl}=\mathbb{M}_{jikl}=\mathbb{M}_{ijlk}=\mathbb{M}_{klij}\\
\tag{$i,j=1,2,3$}
&\alpha_{ij}=\alpha_{ji}
\end{align*}
The inverse of the generalized Hooke's law \eqref{constitutive} is given by~\cite{cheng-1997}
\begin{align}
\label{invconstitutive}
\boldsymbol{\epsilon}=\mathbb{C}\boldsymbol{\sigma}+\frac{1}{3}C\mathbf{B}p
\end{align}
where $\mathbb{C}$ is the fourth order anisotropic compliance tensor, $C(>0)$ is a generalized Hooke's law constant and $\mathbf{B}$ is a generalization of the Skempton pore pressure coefficient $B$~\cite{skempton-1954} for anisotropic poroelasticity. The symmetry of the stress and strain tensor shows that
\begin{align*}
\tag{$i,j,k,l=1,2,3$}
&\mathbb{C}_{ijkl}=\mathbb{C}_{jikl}=\mathbb{C}_{ijlk}=\mathbb{C}_{klij}\\
\tag{$i,j=1,2,3$}
&B_{ij}=B_{ji}
\end{align*}
\subsection{Fluid content}
The fluid content $\zeta$ is given by~\cite{cheng-1997}
\begin{align}
\label{fluidcontent}
\zeta=Cp+\frac{1}{3}C\mathbf{B}:\boldsymbol{\sigma}\equiv \frac{1}{M}p+\boldsymbol{\alpha}:\boldsymbol{\epsilon}
\end{align}
where $M(>0)$ is a generalization of the Biot modulus~\cite{biot3} for anisotropic poroelasticity.
\subsection{The poroelastic tensorial constitutive equation and the strain energy density}
The generalized Hooke's law \eqref{constitutive} and its inverse \eqref{invconstitutive} written in indicial notation as
\begin{align*}
&\sigma_{ij}=\mathbb{M}_{ijkl}\epsilon_{kl}-\alpha_{ij}p\\
&\epsilon_{ij}=\mathbb{C}_{ijkl}\sigma_{ij}+\frac{1}{3}CB_{ij}p
\end{align*}
are rewritten (respectively) in the contracted notation as
\begin{align*}
&\sigma_{\beta}=\mathbb{M}_{\beta\theta}\epsilon_{\gamma}-\alpha_{\beta}p\\
&\epsilon_{\beta}=\mathbb{C}_{\beta \theta}\sigma_{\beta}+\frac{1}{3}CB_{\beta}p
\end{align*}
where the transformation is accomplished by replacing the subscripts $ij$ (or $kl$) by $\beta$ (or $\theta$) using the following rules
\begin{align*}
\begin{array}{ccc|ccc}
ij\,\,(\mathrm{or}\,\,kl) & \longleftrightarrow & \beta\,\,(\mathrm{or}\,\,\theta) & ij\,\,(\mathrm{or}\,\,kl) & \longleftrightarrow & \beta\,\,(\mathrm{or}\,\,\theta)\\
11 & \longleftrightarrow & 1 & 23\,\,(\mathrm{or}\,\,32) & \longleftrightarrow & 4\\
22 & \longleftrightarrow & 2 & 31\,\,(\mathrm{or}\,\,13) & \longleftrightarrow & 5\\
33 & \longleftrightarrow & 3 & 12\,\,(\mathrm{or}\,\,21) & \longleftrightarrow & 6
\end{array}
\end{align*}
and the fourth order tensors $\mathbb{M}_{3\times 3\times 3\times 3}$ and $\mathbb{C}_{3\times 3\times 3\times 3}$ are represented as second order tensors $\mathbb{M}_{6\times 6}$ and $\mathbb{C}_{6\times 6}$ respectively. The constitutive equation for the poromechanics model, under the assumption of zero in-situ stress ($\boldsymbol{\sigma}_0(\mathbf{x})=\mathbf{0}\,\,\forall \mathbf{x}\in \Omega$) and zero in-situ pore pressure ($p_0(\mathbf{x})=0\,\,\forall \mathbf{x}\in \Omega$), is then given by
\begin{align}
\label{cons1}
\boldsymbol{\sigma}_{6\times 1}=\mathbb{M}_{6\times 6}\boldsymbol{\epsilon}_{6\times 1}
-\boldsymbol{\alpha}_{6\times 1}p_{1\times 1}
\end{align}
On the other hand, the constitutive equation for the flow model, under the assumption of zero in-situ stress ($\boldsymbol{\sigma}_0(\mathbf{x})=\mathbf{0}\,\,\forall \mathbf{x}\in \Omega$) and zero in-situ pore pressure ($p_0(\mathbf{x})=0\,\,\forall \mathbf{x}\in \Omega$), is then given by
\begin{align}
\label{cons2}
\zeta_{1\times 1}=\boldsymbol{\alpha}_{6\times 1}:\boldsymbol{\epsilon}_{6\times 1}+\frac{1}{M}p_{1\times 1}
\end{align}
In lieu of \eqref{cons1} and \eqref{cons2}, we can write
\begin{align}
\label{consmatrix}
\overbrace{\left\{\begin{array}{c}
\boldsymbol{\sigma}_{6\times 1}\\
\zeta_{1\times 1}
\end{array}\right\}}^{\boldsymbol{\kappa}_{7\times 1}}=
\overbrace{\begin{bmatrix}
\mathbb{M}_{6\times 6} & -\boldsymbol{\alpha}_{6\times 1}\\
\boldsymbol{\alpha}^T_{6\times 1} & \frac{1}{M}
\end{bmatrix}}^{\mathbb{A}_{7\times 7}}
\overbrace{\left\{\begin{array}{c}
\boldsymbol{\epsilon}_{6\times 1}\\
p_{1\times 1}
\end{array}\right\}}^{\boldsymbol{\gamma}_{7 \times 1}}
\end{align}
\cite{biot-energy} defined the strain energy density of porous elastic medium as
\begin{align}
\label{strainenergy}
\mathscr{U}(\boldsymbol{\upsilon})=\frac{1}{2}(\boldsymbol{\sigma}:\boldsymbol{\epsilon}+p\zeta)
\end{align}
where $\boldsymbol{\upsilon}\equiv (\mathbf{u},p)$ is the poroelastic field. In lieu of \eqref{consmatrix} and \eqref{strainenergy}, the strain energy density of porous elastic medium is given as
\begin{align}
\label{sed}
\mathscr{U}(\boldsymbol{\upsilon})=\frac{1}{2}\boldsymbol{\kappa}^T\boldsymbol{\gamma}
\end{align}
\section{The argument on bounds on effective moduli in linear elasticity using displacement boundary conditions vis-\'a-vis traction boundary conditions}
The language used in this module is similar to the language used in \cite{hollister} due of its ease of explanation. The reader is also refered to \cite{fung} and \cite{zohdi} for a rendition of the classical extremum principles in elasticity, including the principle of minimum strain energy and the principle of minimum complementary energy. Consider the case in a linear elasticity boundary value problem where the in-situ boundary conditions differ from the applied boundary
conditions, but produce the same average RVE strain. In this case the average stiffness predicted
by the RVE analysis must be greater than the actual stiffness by the principle of minimum strain
energy. The in-situ boundary conditions would minimize the energy while the assumed boundary
conditions would be admissible and by definition produce greater energy. The average stress within
the RVE under assumed boundary conditions must be higher to produce a higher energy. The
same argument holds for applied tractions boundary conditions with the principle of minimum
complementary energy. In this situation, the applied traction boundary condition will
produce a higher complementary energy than an in-situ traction condition for the same average
stress giving a higher compliance and therefore a lower stiffness. Thus, RVE analyses under applied
displacements give an upper bound on apparent stiffness while applied tractions give a lower
bound. The expressions for average stress and strain are given in \ref{avgthm}.
\subsection{Homogeneous displacement boundary conditions}
A homogeneous displacement boundary condition is refered to the case of boundary prescribed displacements (over the entire boundary) corresponding to uniform strain (over the entire boundary) as follows 
\begin{align}
\label{hombc1}
u_i(S)=\epsilon_{ij}^0 x_j
\end{align}
Substituting \eqref{hombc1} in \eqref{finalstrainavg}, we get
\begin{align}
\label{blah}
\bar{\epsilon}_{ij}&=\frac{1}{2V}\int\limits_{S} (\epsilon_{ik}^0 x_k n_j+\epsilon_{jk}^0 x_k n_i)=\frac{1}{2V}\int\limits_{V} (\epsilon_{ik}^0 x_{k,j}+\epsilon_{jk}^0 x_{k,i})\\
\nonumber
&=\frac{1}{2V}\int\limits_{V} (\epsilon_{ik}^0 \delta_{kj}+\epsilon_{jk}^0 \delta_{ki})=\frac{1}{2V}\int\limits_{V} (\epsilon_{ij}^0+\epsilon_{ji}^0)\\
\label{clearhai1}
&=\epsilon_{ij}^0
\end{align}
where the second equality in \eqref{blah} follows from the divergence theorem. \eqref{clearhai1} clearly shows that the RVE average strain in case of homogeneous displacement boundary condition is equal to the uniform boundary strain that the boundary condition corresponds to.
\subsection{Homogeneous traction boundary conditions}
A homogeneous traction boundary condition is refered to the case of boundary prescribed tractions (over the entire boundary) corresponding to uniform stress (over the entire boundary) as follows 
\begin{align}
\label{hombc2}
t_i(S)=\sigma_{ij}^0 n_j
\end{align}
Substituting \eqref{hombc2} in \eqref{finalstressavg}, with zero body forces, we get
\begin{align}
\label{random1}
\bar{\sigma}_{ij}&=\frac{1}{2V}\int\limits_{S} (x_j \sigma_{ik}^0 n_k+x_i \sigma_{jk}^0 n_k)=\frac{1}{2V}\int\limits_{V} (x_{j,k} \sigma_{ik}^0 +x_{i,k} \sigma_{jk}^0)\\
\nonumber
&=\frac{1}{2V}\int\limits_{V} (\delta_{jk} \sigma_{ik}^0 +\delta_{ik} \sigma_{jk}^0)=\frac{1}{2V}\int\limits_{V} (\sigma_{ji}^0 +\sigma_{ij}^0)\\
\label{clearhai2}
&=\sigma_{ij}^0
\end{align}
where the second equality in \eqref{random1} follows from the divergence theorem. \eqref{clearhai2} clearly shows that the RVE average stress in case of homogeneous traction boundary condition is equal to the uniform boundary stress that the boundary condition corresponds to. 
\section{Procedural framework to arrive at bounds on effective poroelastic moduli}
We now invoke the principles derived in Section 3 for linear elastic solids to arrive at bounds on effective moduli for heterogeneous linear poroelastic solids. The homogeneous displacement boundary condition applicable to linear elasticity is replaced by a homogeneous displacement-pressure boundary condition. The homogeneous traction boundary condition applicable to linear elasticity is replaced by a homogeneous traction-fluid content boundary condition.
\subsection{Homogeneous displacement-pressure boundary condition}
Let the poroelastic body with no body forces be subjected to the homogeneous boundary condition
\begin{equation}
\label{hombc}
\left.\begin{array}{c}
u_i(\partial \Omega)=\epsilon_{ij}^0x_j\\
p(\partial \Omega)=p^0
\end{array}\right\}
\end{equation}
The $\boldsymbol{\gamma}^0$ vector in \eqref{consmatrix} can be separated into 6 vectors in each of which there occurs only one non-vanishing strain $\epsilon_{ij}^0$ and one vector in which there occurs only one non-vanishing pressure $p^0$ as follows
\begin{align}
\label{each}
\left\{\hspace{-6pt}\begin{array}{c}
\gamma_1^0\\
\gamma_2^0\\
\gamma_3^0\\
\gamma_4^0\\
\gamma_5^0\\
\gamma_6^0\\
\gamma_7^0
\end{array}\hspace{-6pt}\right\}
=\left\{\hspace{-6pt}\begin{array}{c}
\gamma_1^0\\
0\\
0\\
0\\
0\\
0\\
0
\end{array}\hspace{-6pt}\right\}+
\left\{\hspace{-6pt}\begin{array}{c}
0\\
\gamma_2^0\\
0\\
0\\
0\\
0\\
0
\end{array}\hspace{-6pt}\right\}+
\left\{\hspace{-6pt}\begin{array}{c}
0\\
0\\
\gamma_3^0\\
0\\
0\\
0\\
0
\end{array}\hspace{-6pt}\right\}+
\left\{\hspace{-6pt}\begin{array}{c}
0\\
0\\
0\\
\gamma_4^0\\
0\\
0\\
0
\end{array}\hspace{-6pt}\right\}+
\left\{\hspace{-6pt}\begin{array}{c}
0\\
0\\
0\\
0\\
\gamma_5^0\\
0\\
0
\end{array}\hspace{-6pt}\right\}+
\left\{\hspace{-6pt}\begin{array}{c}
0\\
0\\
0\\
0\\
0\\
\gamma_6^0\\
0
\end{array}\hspace{-6pt}\right\}+
\left\{\hspace{-6pt}\begin{array}{c}
0\\
0\\
0\\
0\\
0\\
0\\
\gamma_7^0
\end{array}\hspace{-6pt}\right\}
\end{align}
In lieu of \eqref{each}, the poroelastic boundary field $(\mathbf{u},p)$ can be decomposed as
\begin{align}
\label{eachagain}
\left\{\hspace{-6pt}\begin{array}{c}
u_1\\
u_2\\
u_3\\
p
\end{array}\hspace{-6pt}\right\}&=
\left\{\hspace{-6pt}\begin{array}{c}
\gamma_1^0x_1\\
0\\
0\\
0
\end{array}\hspace{-6pt}\right\}+
\left\{\hspace{-6pt}\begin{array}{c}
0\\
\gamma_2^0x_2\\
0\\
0
\end{array}\hspace{-6pt}\right\}+
\left\{\hspace{-6pt}\begin{array}{c}
0\\
0\\
\gamma_3^0x_3\\
0
\end{array}\hspace{-6pt}\right\}
+\left\{\hspace{-6pt}\begin{array}{c}
\gamma_4^0x_3\\
0\\
\gamma_4^0x_1\\
0
\end{array}\hspace{-6pt}\right\}+
\left\{\hspace{-6pt}\begin{array}{c}
0\\
\gamma_5^0x_3\\
\gamma_5^0x_2\\
0
\end{array}\hspace{-6pt}\right\}+
\left\{\hspace{-6pt}\begin{array}{c}
\gamma_6^0x_2\\
\gamma_6^0x_1\\
0\\
0
\end{array}\hspace{-6pt}\right\}+
\left\{\hspace{-6pt}\begin{array}{c}
0\\
0\\
0\\
\gamma_7^0
\end{array}\hspace{-6pt}\right\}
\end{align}
Because of the superposition principle of the linear theory of elasticity (and hence by natural extension poroelasticity), the poroelastic $\boldsymbol{\upsilon}\equiv (\mathbf{u},p)$ field which is produced by \eqref{hombc} is equal to the sum of the seven fields which are produced by the application of each of \eqref{eachagain}, separately, on the boundary. Suppose that $\gamma_1^0=1$ and let the resulting poroelastic $\boldsymbol{\upsilon}\equiv (\mathbf{u},p)$ field be denoted by $\boldsymbol{\upsilon}^{(\gamma_1)}\equiv \{\upsilon_1^{(\gamma_1)},v_2^{(\gamma_1)},v_3^{(\gamma_1)},v_4^{(\gamma_1)}\}^T$. Then, when $\gamma_1^0\neq 1$, the field is by linearity $\gamma_1^0\boldsymbol{\upsilon}^{(\gamma_1)}$. Similar considerations for each of $\gamma_2^0$, $\gamma_3^0$, $\gamma_4^0$, $\gamma_5^0$, $\gamma_6^0$ and $\gamma_7^0$ on the right hand side of \eqref{eachagain} and superposition show that the poroelastic $\boldsymbol{\upsilon}\equiv(\mathbf{u},p)$ field in the body due to \eqref{hombc} on the boundary can be written in the form 
\begin{equation}
\label{badass}
\boldsymbol{\upsilon}=\gamma_k^0 \boldsymbol{\upsilon}^{(\gamma_k)}
\end{equation}
where $k$ is summed. The $\boldsymbol{\gamma}$ field at any point is then given by
\begin{align}
\label{gamma}
\boldsymbol{\gamma}(\mathbf{x})=\gamma_k^0 \boldsymbol{\gamma}\big(\boldsymbol{\upsilon}^{(\gamma_k)}\big)
\end{align}
Finally, the $\boldsymbol{\kappa}$ at any point is given in view of \eqref{consmatrix} and \eqref{gamma} as
\begin{align}
\label{final}
\boldsymbol{\kappa}(\mathbf{x})=\gamma_k^0\mathbb{A}(\mathbf{x}) \boldsymbol{\gamma}\big(\boldsymbol{\upsilon}^{(\gamma_k)}\big)
\end{align}
where $\mathbb{A}(\mathbf{x})$ are the space dependent poroelastic moduli of the heterogeneous body and $k$ is summed. Let \eqref{final} be volume averaged over a RVE. The result is written in the form
\begin{align*}
\bar{\boldsymbol{\kappa}}=\mathbb{A}^*\gamma_k^0
\end{align*}
where the upper bound on effective poroelastic moduli is obtained as
\begin{align*}
\tag{upper bound}
\boxed{\mathbb{A}^*\equiv \frac{1}{V}\int\limits_V \mathbb{A}(\mathbf{x})\boldsymbol{\gamma}\big(\boldsymbol{\upsilon}^{(\gamma_k)}\big)}
\end{align*}
\subsection{Homogeneous traction-fluid content boundary condition}
Let the poroelastic body with no body forces be subjected to the homogeneous boundary condition
\begin{equation}
\label{hombcfirse}
\left.\begin{array}{c}
t_i(\partial \Omega)=\sigma_{ij}^0n_j\\
\zeta(\partial \Omega)=\zeta^0
\end{array}\right\}
\end{equation}
The $\boldsymbol{\kappa}^0$ vector in \eqref{consmatrix} can be separated into 6 vectors in each of which there occurs only one non-vanishing stress $\sigma_{ij}^0$ and one vector in which there occurs only one non-vanishing fluid content $\zeta^0$ as follows
\begin{align}
\label{eachfirse}
\left\{\hspace{-6pt}\begin{array}{c}
\kappa_1^0\\
\kappa_2^0\\
\kappa_3^0\\
\kappa_4^0\\
\kappa_5^0\\
\kappa_6^0\\
\kappa_7^0
\end{array}\hspace{-6pt}\right\}
=\left\{\hspace{-6pt}\begin{array}{c}
\kappa_1^0\\
0\\
0\\
0\\
0\\
0\\
0
\end{array}\hspace{-6pt}\right\}+
\left\{\hspace{-6pt}\begin{array}{c}
0\\
\kappa_2^0\\
0\\
0\\
0\\
0\\
0
\end{array}\hspace{-6pt}\right\}+
\left\{\hspace{-6pt}\begin{array}{c}
0\\
0\\
\kappa_3^0\\
0\\
0\\
0\\
0
\end{array}\hspace{-6pt}\right\}+
\left\{\hspace{-6pt}\begin{array}{c}
0\\
0\\
0\\
\kappa_4^0\\
0\\
0\\
0
\end{array}\hspace{-6pt}\right\}+
\left\{\hspace{-6pt}\begin{array}{c}
0\\
0\\
0\\
0\\
\kappa_5^0\\
0\\
0
\end{array}\hspace{-6pt}\right\}+
\left\{\hspace{-6pt}\begin{array}{c}
0\\
0\\
0\\
0\\
0\\
\kappa_6^0\\
0
\end{array}\hspace{-6pt}\right\}+
\left\{\hspace{-6pt}\begin{array}{c}
0\\
0\\
0\\
0\\
0\\
0\\
\kappa_7^0
\end{array}\hspace{-6pt}\right\}
\end{align}
In lieu of \eqref{eachfirse}, the poroelastic boundary field $(\mathbf{t},\zeta)$ can be decomposed as
\begin{align}
\label{eachagainfirse}
\left\{\hspace{-6pt}\begin{array}{c}
t_1\\
t_2\\
t_3\\
\zeta
\end{array}\hspace{-6pt}\right\}&=
\left\{\hspace{-6pt}\begin{array}{c}
\kappa_1^0n_1\\
0\\
0\\
0
\end{array}\hspace{-6pt}\right\}+
\left\{\hspace{-6pt}\begin{array}{c}
0\\
\kappa_2^0n_2\\
0\\
0
\end{array}\hspace{-6pt}\right\}+
\left\{\hspace{-6pt}\begin{array}{c}
0\\
0\\
\kappa_3^0n_3\\
0
\end{array}\hspace{-6pt}\right\}
+\left\{\hspace{-6pt}\begin{array}{c}
\kappa_4^0n_3\\
0\\
\kappa_4^0n_1\\
0
\end{array}\hspace{-6pt}\right\}+
\left\{\hspace{-6pt}\begin{array}{c}
0\\
\kappa_5^0n_3\\
\kappa_5^0n_2\\
0
\end{array}\hspace{-6pt}\right\}+
\left\{\hspace{-6pt}\begin{array}{c}
\kappa_6^0n_2\\
\kappa_6^0n_1\\
0\\
0
\end{array}\hspace{-6pt}\right\}+
\left\{\hspace{-6pt}\begin{array}{c}
0\\
0\\
0\\
\kappa_7^0
\end{array}\hspace{-6pt}\right\}
\end{align}
Because of the superposition principle of the linear theory of elasticity (and hence by natural extension poroelasticity), the poroelastic $\boldsymbol{\Upsilon}\equiv (\mathbf{t},\zeta)$ field which is produced by \eqref{hombcfirse} is equal to the sum of the seven fields which are produced by the application of each of \eqref{eachagainfirse}, separately, on the boundary. Suppose that $\kappa_1^0=1$ and let the resulting poroelastic $\boldsymbol{\Upsilon}\equiv (\mathbf{t},\zeta)$ field be denoted by $\boldsymbol{\Upsilon}^{(\kappa_1)}\equiv \{\Upsilon_1^{(\kappa_1)},\Upsilon_2^{(\kappa_1)},\Upsilon_3^{(\kappa_1)},\Upsilon_4^{(\kappa_1)}\}^T$. Then, when $\kappa_1^0\neq 1$, the field is by linearity $\kappa_1^0\boldsymbol{\Upsilon}^{(\kappa_1)}$. Similar considerations for each of $\kappa_2^0$, $\kappa_3^0$, $\kappa_4^0$, $\kappa_5^0$, $\kappa_6^0$ and $\kappa_7^0$ on the right hand side of \eqref{eachagainfirse} and superposition show that the poroelastic $\boldsymbol{\Upsilon}\equiv(\mathbf{t},\zeta)$ field in the body due to \eqref{hombcfirse} on the boundary can be written in the form 
\begin{equation}
\label{badassfirse}
\boldsymbol{\Upsilon}=\kappa_k^0 \boldsymbol{\Upsilon}^{(\kappa_k)}
\end{equation}
where $k$ is summed. The $\boldsymbol{\kappa}$ field at any point is then given by
\begin{align}
\label{gammafirse}
\boldsymbol{\kappa}(\mathbf{x})=\kappa_k^0 \boldsymbol{\kappa}\big(\boldsymbol{\Upsilon}^{(\kappa_k)}\big)
\end{align}
Finally, the $\boldsymbol{\gamma}$ at any point is given in view of \eqref{consmatrix} and \eqref{gammafirse} as
\begin{align}
\label{finalfirse}
\boldsymbol{\gamma}(\mathbf{x})=\kappa_k^0\mathbb{A}^{-1}(\mathbf{x}) \boldsymbol{\kappa}\big(\boldsymbol{\Upsilon}^{(\kappa_k)}\big)
\end{align}
where $\mathbb{A}(\mathbf{x})$ are the space dependent poroelastic moduli of the heterogeneous body and $k$ is summed. Let \eqref{finalfirse} be volume averaged over a RVE. The result is written in the form
\begin{align*}
\bar{\boldsymbol{\gamma}}=\mathbb{A}^{*^{-1}}\kappa_k^0
\end{align*}
where the effective poroelastic compliance is obtained as
\begin{align*}
\mathbb{A}^{*^{-1}}\equiv \frac{1}{V}\int\limits_V \mathbb{A}^{-1}(\mathbf{x})\boldsymbol{\kappa}\big(\boldsymbol{\Upsilon}^{(\kappa_k)}\big)
\end{align*}
which is inverted to obtain the lower bound on the effective poroelastic moduli
\begin{align*}
\tag{lower bound}
\boxed{\mathbb{A}^*\equiv \bigg(\frac{1}{V}\int\limits_V \mathbb{A}^{-1}(\mathbf{x})\boldsymbol{\kappa}\big(\boldsymbol{\Upsilon}^{(\kappa_k)}\big)\bigg)^{-1}}
\end{align*}
\subsection{Invoking statistical homogeneity}
The essence of statistical homogeneity is explained in \ref{essence}. The crux of the argument is that the stress and strain fields in a very large heterogeneous body that is statistically homogeneous, subject to homogeneous boundary conditions, is statistically homogeneous. The argument is that if the field is statistically homogeneous, the volume average taken over the RVE approaches the whole body average, wherever the RVE may be located. The reader is refered to \cite{hashinnasa} for a finer read on the concept of statistical homogeneity.
\section{Conclusions and outlook}
The upper and lower bounds on effective poroelastic moduli have been derived. This is of significant advantage for highly heterogeneous structures since resolution of fine scale features can be computationally expensive with any discretization technique. The derivation is bourne out of the concepts of RVE, statistical homogeneity and superposition principle in poroelastic solids.
\appendix
\section{Averaging theorems}\label{avgthm}
\begin{figure}[h]
\centering
\includegraphics[scale=2.5]{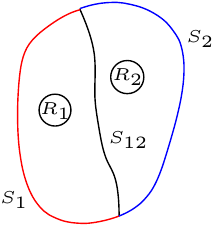}
\caption{Two phase body. $S_1\cup S_2\equiv S\oplus S_{12}\oplus S_{12}$}
\label{twophase}
\end{figure}
As shown in Figure \ref{twophase}, consider a two phase body with phases occupying regions $R_1$ and $R_2$ with displacement fields $u_i^{(1)}(\mathbf{x})$ and $u_i^{(2)}(\mathbf{x})$ respectively. The volume of the two phase body is $V$, the phase volumes are $V_1$ and $V_2$, the bounding surface is $S$ and the interface is $S_{12}$. 
\subsection{Expression for average strain}
The strains $\epsilon_{ij}$ are defined as
\begin{align}
\label{ss}
\epsilon_{ij}=\frac{1}{2}(u_{i,j}(\mathbf{x})+u_{j,i}(\mathbf{x}))
\end{align}
The volume average $\bar{\epsilon}_{ij}$ of $\epsilon_{ij}$ is given by
\begin{align}
\label{ss1}
\bar{\epsilon}_{ij}=\frac{1}{V}\int\limits_V \epsilon_{ij}(\mathbf{x})
\end{align}
Substituting \eqref{ss} in \eqref{ss1}, we get
\begin{align*}
\bar{\epsilon}_{ij}=\frac{1}{2V}\int\limits_V (u_{i,j}(\mathbf{x})+u_{j,i}(\mathbf{x}))
\end{align*}
which, in lieu of the divergence theorem, can be written as
\begin{align}
\label{ss2}
\bar{\epsilon}_{ij}=\frac{1}{2V}\bigg[\int\limits_{S_1} (u_i^{(1)}n_j+u_j^{(1)}n_i)+\int\limits_{S_2} (u_i^{(2)}n_j+u_j^{(2)}n_i)\bigg]
\end{align}
where $S_1$ and $S_2$ are the bounding surfaces of $R_1$ and $R_2$. Now each of $S_1$ and $S_2$ are composed of part of the external surface $S$ and the interface $S_{12}$. Therefore, \eqref{ss2} may be rewritten as
\begin{align}
\label{ss3}
\bar{\epsilon}_{ij}&=\frac{1}{2V}\bigg[\int\limits_{S} (u_i n_j+u_j n_i)+\overbrace{\int\limits_{S_{12}} (u_i^{(1)}n_j+u_j^{(1)}n_i)+\int\limits_{S_{12}} (u_i^{(2)}n_j+u_j^{(2)}n_i)}^0\bigg]\\
\label{finalstrainavg}
&=\frac{1}{2V}\int\limits_{S} (u_i n_j+u_j n_i)
\end{align}
where the interface integral terms in \eqref{ss3} cancel out as $n_i$ (and $n_j$) is always the outward normal.
\subsection{Expression for average stress}
Let the stress field inside the body be $\sigma_{ij}(\mathbf{x})$ and the body force per unit volume be $f_i(\mathbf{x})$. The body is assumed to be in quasi-static equilibrium, so that at every point,
\begin{align*}
\sigma_{ij,j}+f_i=0
\end{align*}
On the external surface $S$, the tractions $t_i(S)=\sigma_{ij}n_j$ are prescribed. The average stress is defined by
\begin{align}
\label{avgstress}
\bar{\sigma}_{ij}=\frac{1}{V}\int\limits_V \sigma_{ij}
\end{align}
We substitute the following
\begin{align*}
(\sigma_{ik}x_j)_{,k}=\sigma_{ik,k}x_j+\sigma_{ik}\delta_{jk}=-f_ix_j+\sigma_{ij}
\end{align*}
into \eqref{avgstress}, to get
\begin{align}
\label{mama}
\bar{\sigma}_{ij}&=\frac{1}{V}\int\limits_V ((\sigma_{ik}x_j)_{,k}+f_ix_j)=\frac{1}{V}\bigg[\int\limits_S \sigma_{ik}x_j n_k+\int\limits_V f_ix_j\bigg]\\
\nonumber
&=\frac{1}{V}\bigg[\int\limits_{S_1} x_j \sigma_{ik}^{(1)}n_k^{(1)}+\int\limits_{S_2} x_j \sigma_{ik}^{(2)}n_k^{(2)}+\int\limits_V f_ix_j\bigg]\\
\label{mama1}
&=\frac{1}{V}\bigg[\int\limits_{S} x_j \sigma_{ik}n_k+\overbrace{\int\limits_{S_{12}} x_j \overbrace{\sigma_{ik}^{(1)}n_k^{(1)}}^{t_i^{(1)}}+\int\limits_{S_{12}} x_j \overbrace{\sigma_{ik}^{(2)}n_k^{(2)}}^{t_i^{(2)}}}^0+\int\limits_V f_ix_j\bigg]
\end{align}
where the second equality in \eqref{mama} with the surface integral follows from the divergence theorem and the integrands of the interface surface integrals in \eqref{mama1} cancel one another at each interface point and thus the two surface integrals cancel each other. Also, since $\sigma_{ij}(\mathbf{x})=\sigma_{ji}(\mathbf{x})$, it follows that $\bar{\sigma}_{ij}=\bar{\sigma}_{ji}$. Thus \eqref{mama1} can be symmetrized in the form
\begin{align}
\label{finalstressavg}
\bar{\sigma}_{ij}=\frac{1}{2V}\bigg[\int\limits_{S} (x_j t_i+x_i t_j)+\int\limits_V (x_jf_i+x_if_j)\bigg]
\end{align}
\section{Statistical homogeneity}\label{essence}
Consider a collection of $N$ fibrous cylindrical specimens, each of which is refered to a system of axes, the origin of which is at the same point in each specimen. The specimens have the same external geometry, however, their phase geometries, i.e. internal geometries may be quite different. In the theory of random processes, such a collection of specimens is called an ensemble and each specimen is a member of the ensemble. We consider the two points $B_1$, $B_2$ in each specimen member of the ensemble, which have the same position vectors $\mathbf{x}^1$ and $\mathbf{x}^2$ in each specimen. The probability that $B_1$ is in $R_1$ and $B_2$ is in $R_1$ is
\begin{align*}
P(B_1\supset R_1,B_2\supset R_1)=P_{11}(\mathbf{x}^1,\mathbf{x}^2)=\lim\limits_{N\rightarrow \infty}\frac{N_{11}}{N}
\end{align*}
Similarly, we have
\begin{align*}
&P(B_1\supset R_1,B_2\supset R_2)=P_{12}(\mathbf{x}^1,\mathbf{x}^2)=\lim\limits_{N\rightarrow \infty}\frac{N_{12}}{N}\\
&P(B_1\supset R_2,B_2\supset R_1)=P_{21}(\mathbf{x}^1,\mathbf{x}^2)=\lim\limits_{N\rightarrow \infty}\frac{N_{21}}{N}\\
&P(B_1\supset R_2,B_2\supset R_2)=P_{22}(\mathbf{x}^1,\mathbf{x}^2)=\lim\limits_{N\rightarrow \infty}\frac{N_{22}}{N}
\end{align*}
Here $N_{11}$ is the number of times that both points fall simultaneously into phase $1$, with analogous interpretations for $N_{12}$, $N_{21}$ and $N_{22}$. There are similar eight three point probabilities and in general $2^n$, $n$ point probabilities. Such $n$ point probabilities may be written in the form 
\begin{align}
\label{boom}
P_{i_1,i_2,..,i_n}(\mathbf{x}^1,\mathbf{x}^2,..,\mathbf{x}^n)
\end{align}
where each of the subscripts $i_1$, $i_2$, ..., $i_n$ assumes either one of the phase numbers $1$, $2$ and its position in the subscript sequence is attached to that corresponding to the position vector within the parenthesis. We now finally proceed to define the concept of statistical homogeneity. For this purpose, the system of $n$ points entering into \eqref{boom} may be considered as a rigid body which is described by the vector differences
\begin{align}
\label{boomshaka}
\mathbf{r}^1=\mathbf{x}^1-\mathbf{x}^n,\,\,\mathbf{r}^2=\mathbf{x}^2-\mathbf{x}^n,\cdots,\mathbf{x}^{n-1}=\mathbf{x}^{n-1}-\mathbf{x}^n
\end{align}
Suppose that any multipoint probability such as \eqref{boom} depends only on the relative configuration of the points and not on their absolute position with respect to the coordinate system; then the ensemble is called statistically homogeneous. Mathematically this means
\begin{align}
\label{stathom}
P_{i_1},P_{i_2}\cdots i_n(\mathbf{x}^1,\mathbf{x}^2,..,\mathbf{x}^{n})=P_{i_1,i_2,..,i_n}(\mathbf{r}^1,\mathbf{r}^2,..,\mathbf{r}^{n-1})
\end{align}
\bibliographystyle{unsrt} 
\bibliography{diss1}
\end{document}